%% file: SC_Mag_v12.tex
\documentclass[article]{IEEEtran}
\input{preamble3.tex}
\usepackage{tabularx}
\usepackage{graphics}
\usepackage{bbm}

\title{Symbiotic Communications: Where Marconi Meets Darwin}

\author{Ying-Chang Liang, \emph{Fellow, IEEE}, Ruizhe Long, Qianqian Zhang, and Dusit Niyato, \emph{Fellow, IEEE}\\

\thanks{

This work has been submitted to the IEEE for possible publication. Copyright may be transferred without notice, after which this version may no longer be accessible.

Y.-C. Liang is with the Center for Intelligent Networking and Communications (CINC), University of Electronic Science and Technology of China (UESTC), Chengdu 611731, China, and also with Shenzhen Institute of Advanced Studies, UESTC, Shenzhen, China (e-mail: liangyc@ieee.org).

R. Long and Q.~Zhang are with the National Key Laboratory of Science and Technology on Communications, and the Center for Intelligent Networking and Communications (CINC), University of Electronic Science and Technology of China (UESTC), Chengdu 611731, China (e-mails:
ruizhelong@gmail.com; qqzhang\_kite@163.com).

D. Niyato is with the School of Computer Science and Engineering, Nanyang Technological University, Singapore 639798 (e-mail: dniyato@ntu.edu.sg).
}}
\begin{document}
\maketitle
\begin{abstract}
With the proliferation of wireless applications, the electromagnetic (EM) space is becoming more and more crowded and complex. This makes it a challenging task to accommodate the growing number of radio systems with limited radio resources. In this paper, by considering the EM space as a radio ecosystem, and leveraging the analogy to the natural ecosystem in biology, a novel symbiotic communication (SC) paradigm is proposed through which the relevant radio systems, called symbiotic radios (SRs), in a radio ecosystem form a symbiotic relationship (e.g., mutualistic symbiosis) through intelligent resource/service exchange. Radio resources include, e.g., spectrum, energy, and infrastructure, while typical radio services are communicating, relaying, and computing. The symbiotic relationship can be realized via either symbiotic coevolution or symbiotic synthesis. In symbiotic coevolution, each SR is empowered with an evolutionary cycle alongside the multi-agent learning, while in symbiotic synthesis, the SRs ingeniously optimize their operating parameters and transmission protocols by solving a multi-objective optimization problem. Promisingly, the proposed SC paradigm breaks the boundary of radio systems, thus providing us a fresh perspective on radio resource management and new guidelines to design future wireless communication systems.
\end{abstract}

\section{Introduction}


Since Guglielmo Marconi's breakthrough radio experiment in 1895, radio technology has evolved from simple radio telephony to complex systems such as broadcasting, radar, and wireless communication systems. Marconi may never imagine that his invention could become such an indispensable technology to the modern society. This is evident by the fact that, nowadays, for any given area, we are surrounded by various wireless systems, such as WiFi, digital TV and cellular networks, and each system may be further required to support versatile applications~\cite{tse2005fundamentals}.
With the proliferation of wireless applications, the electromagnetic (EM) space is becoming more complex than ever before, making it a challenging task to accommodate the growing number of radio systems with the limited radio resources.


Traditionally, to avoid inter-system interference, each radio system is assigned with a set of dedicated radio resources, such as radio spectrum, and much effort has been spent to improve the utilization efficiency for the assigned resources. Using radio spectrum as an example, the representative technologies include, e.g., multiple-input multiple-output (MIMO) and ultra dense network (UDN). MIMO exploits multiple antennas to realize enhanced spectrum efficiency through spatial multiplexing, while UDN improves the coverage and capacity of the system through deploying dense networks. With more and more radio systems being introduced into the EM space, the radio spectrum is becoming severely scarce. One way to make use of the spectrum resource more efficiently is cognitive radio (CR)~\cite{mitola1999cognitive}, which allows two or more radio systems to share the same spectrum~\cite{haykin2005cognitive}. However, the secondary radio system in CR could cause non-avoidable interference to the primary system, though such interference is controllable within an acceptable level~\cite{liang2020dynamic}. The interfering nature of conventional spectrum sharing prevents the wide adoption of CR in practical applications so far.


Recently, mutualistic spectrum sharing has been observed in cognitive backscatter communication (CBC) systems~\cite{yang2018cooperative}, in which the active transmission achieves multipath diversity with the help of the backscatter transmission, while the backscatter device gains the transmission opportunity using the RF source and radio spectrum of the active transmission. Such mutualism advocates the coexistence of cellular and internet-of-things (IoT) networks in the same spectrum~\cite{liang2020symbiotic}.
Interestingly, the above radio interaction is similar to the mutualistic symbiosis between the roots of the plants and the fungus in biology~\cite{margulis1991symbiosis}. In particular, the fungus colonizes the plant roots and is provided with carbohydrates, sucrose and glucose. In exchange, the plant benefits from the fungi’s higher water and mineral absorption capabilities. Such biological symbiosis motivates us to rethink the design methodology for wireless communication systems to achieve mutual benefits among them.




In this paper, by considering the EM space as a radio ecosystem, and leveraging the analogy to the natural ecosystem in biology, we propose a novel \emph{symbiotic communication (SC) paradigm in which the relevant radio systems, also called symbiotic radios (SRs), in a radio ecosystem form a symbiotic relationship through intelligent resource/service exchange}. Radio resources include, e.g., spectrum, infrastructure, and energy, while typical radio services include communicating, relaying, and computing services. Similar to biological symbiosis, there are different types of symbiotic relationships, including, e.g., parasitism, commensalism, and mutualism, but the most promising one is mutualism, in which all the SRs receive benefits. To achieve, while each SR has its own design objective, in SC, all SRs consider the collective objectives of the SRs and achieve mutual benefits through intelligent resource and service exchange. Thus different resource bottlenecks for each SR can complement each other to support the diversified objectives.

The symbiotic relationship can be realized via either symbiotic coevolution or symbiotic synthesis. In symbiotic coevolution, each SR is empowered with an evolutionary cycle alongside the multi-agent learning, while in symbiotic synthesis, the SRs ingeniously optimize their operating parameters and transmission protocols by solving a multi-objective optimization problem.

The rest of this paper is organized as follows. We start by outlining the biological symbiosis and then give the explicit definition and rationale of SC in Section~\ref{sec:bio-symb} and~\ref{sec:SR}, respectively. Then we investigate two typical SC paradigms in Section~\ref{sec:SC}, obligate SC and facultative SC. We discuss how SC achieves the mutualistic symbiosis via symbiotic coevolution in Section~\ref{sec:exchange}, and establish a multi-objective optimization model to describe symbiotic synthesis in Section~\ref{sec:MOO}. Discussions on technical challenges and research directions in this emerging field are given in Section~\ref{sec:chall}.

\begin{figure}
  \centering
  \includegraphics[width=0.99\columnwidth] {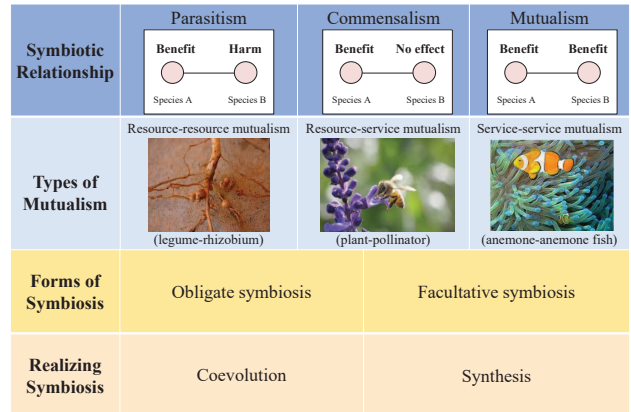}
  \caption{\small Biological symbiosis.}
  \label{fig:SRtobio}
  \end{figure}

\section{Biological Symbiosis} \label{sec:bio-symb}

Human beings live in a natural ecosystem, which accommodates massive plants, animals, and other organisms of diverse species with precious natural resources. In 1859, Charles Darwin published his famous book ``\emph{On the Origin of Species}'', which laid the foundations on evolutionary theory.
In 1960s, biologists found that species's cooperation and interdependence play an important role in the evolutionary process, in addition to competition~\cite{margulis1991symbiosis}. Specifically, multiple symbiotic partners can be of mutual benefits to each other by exchanging resources such as materials, energy, and/or services such as pollination and physical protection, yielding biological symbiosis~\cite{waser2006plant}. Such Neo-Darwinism advocates that the evolutionary process is not an individual behavior of a specific species, but a cooperative behavior of two or more species.

\begin{figure*}[t]
  \centering
  \includegraphics[width=1.99\columnwidth] {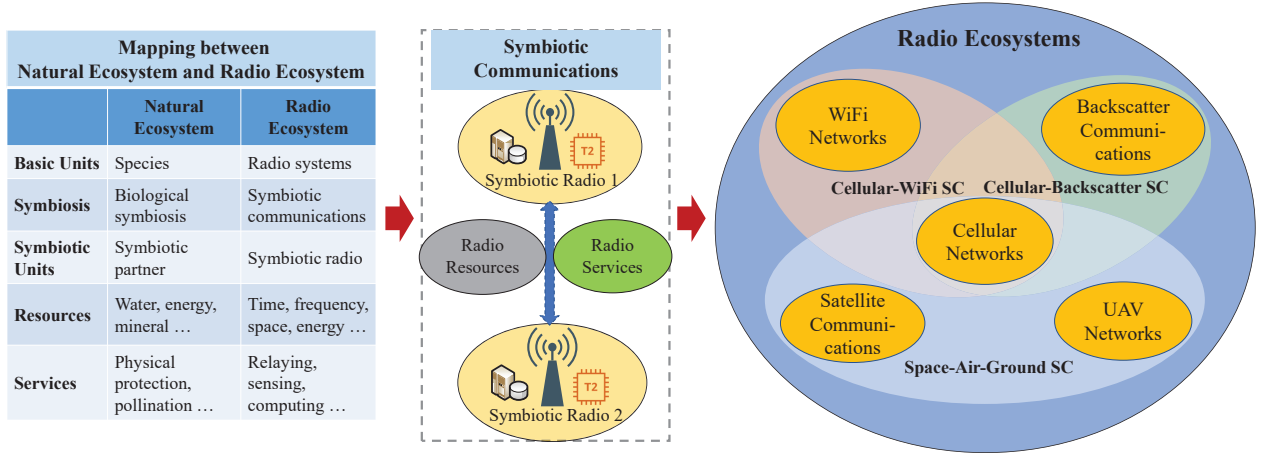}
  \caption{\small Radio ecosystems and symbiotic communications.}
  \label{fig:Radio}
\end{figure*}

In nature, a biological ecosystem is composed of biotic components and abiotic components. The biotic components refer to species like animals and plants, while the abiotic components include physical resources that affect the species in the ecosystem. According to the dependence of the symbiotic partners, there are two types of symbiosis: \emph{obligate symbiosis} and \emph{facultative symbiosis}. In obligate symbiosis, the two close-living organisms are so close to each other that one or both of them cannot survive without the other, while for facultative symbiosis, the two organisms can survive separately but engage in a symbiotic partnership through choice.

From the perspective of mutual benefits, there are different types of \emph{symbiotic relationships}, including parasitism, commensalism, mutualism, and so on. The most promising one is mutualism, for which the symbiotic partners benefit from the interactions. There are three ways to achieve mutualism, according to the resource/service exchanged in the ecosystem.

\subsubsection{Resource-resource mutualism}
Each of the symbiotic partners supplies the nutritional resources needed by the other. One example is the legume-rhizobium mutualism in which the rhizobia colonize the legume plant roots and help the plants to produce the protein, and in return, the plants supply carbohydrates to the rhizobia as the energy source.

\subsubsection{Resource-service mutualism}
Each of the symbiotic partners pays a resource reward for a service. The best-known example is the plant-pollinator mutualism in which the flowers supply the nectar to the bees, while the bees provide the pollination service to the flowers.

\subsubsection{Service-service mutualism}
Each of the two symbiotic partners receives a service from the other. One example is the anemone-anemone fish mutualism in which the sea anemone provides the anemone fish with a shelter and food, and in return, the anemone fish protects the anemone from its predators and parasites.

Finally, biological symbiosis is usually realized through coevolution among the symbiotic partners, which usually takes a long process lasting generations. Biologists have recently proposed symbiotic synthesis which artificially creates the symbiotic relationship between the species without requiring long evolutionary process~\cite{geddes2019engineering}. As a summary, Fig.~\ref{fig:SRtobio} shows the key elements and concepts in biological symbiosis.

\section{Symbiotic Communications} \label{sec:SR}

Analogous to biological ecosystem, a radio ecosystem consists of a set of radio systems together with the radio resources. Radio systems are regarded as the biotic components of the ecosystem which consume the radio resources (e.g., time, frequency, energy and space) to accomplish specific tasks. The range of radio systems is rather wide, including but not limited to broadcasting, radar, and wireless communication systems.

In earlier days, there are abundant radio resources (e.g., spectrum) but few radio systems. Thus, each radio system is allocated with exclusive resources. Modern radio ecosystem however has to accommodate a large number of radio systems due to the proliferation of wireless applications, but the radio resources (especially radio spectrum) are limited, causing interference and competition in the radio ecosystem and leading to resource scarcity problems.


To solve the above problem, the symbiotic communication (SC) paradigm is proposed to enable the relevant radio systems, called symbiotic radios (SRs), to cooperate and realize mutual benefits. Specifically, SC is an intelligent radio ecosystem in which the SRs form a symbiotic relationship through intelligent resource/service exchange. The radio resources include, e.g., time, frequency, space, energy and infrastructure, while the radio services include, e.g., direct communicating, relaying, and computing services.

While each SR has its own design objective in terms of throughput, power consumption, and latency, SC considers the collective objectives of all SRs and support resource and service exchange for the purpose of achieving mutual benefits for all SRs. The rationale behind the SC paradigm is that different resource bottlenecks for each SR can complement each other to support the diversified objectives. This avoids the situations where some SRs exclusively occupy the radio resource and still could not support their applications alone due to stringent requirements, thereby enhancing the overall utilization of the resources.





Similar to biology, in SC, there exist different types of symbiotic relationships among the SRs, including parasitism, commensalism, mutualism, and so on. Also, according to the interdependence of the SRs, we can classify SC into two categories, one is obligate SC, the other is facultative SC. Finally, SC can be realized through symbiotic coevolution in which each SR is empowered with an evolutionary cycle, or through symbiotic synthesis which designs the key operating parameters of each SR to deal with collective objectives.

Fig.~\ref{fig:Radio} summarizes the mapping between biological ecosystem and radio ecosystem, and the key elements and concepts in SC. It is seen that the cellular networks can form SC with various types of networks.

\section{Obligate SC and Facultative SC}\label{sec:SC}

\begin{figure}[t]
  \centering
  \includegraphics[width=1.0\columnwidth] {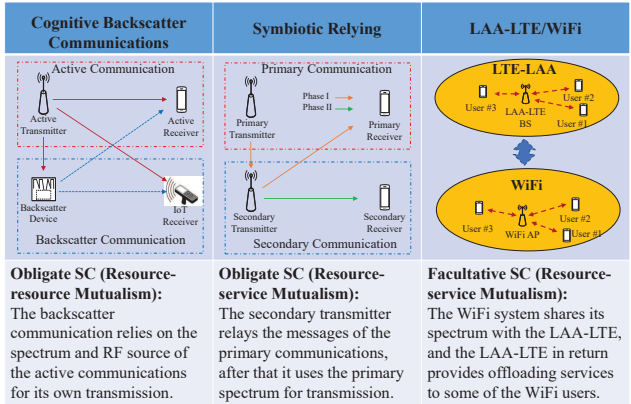}
  \caption{\small Obligate SC: one SR heavily relies on the other.\\
   Facultative SC: the SRs are relatively independent.}
  \label{fig:CBC}
  \vspace{-0.5cm}
\end{figure}

In this section, the definitions of obligate SC and facultative SC are given. Three SC example are also presented to illustrate SRs' dependency and how the radio symbiosis is formed.



\subsection{Obligate SC}

\begin{figure}[t]
  \centering
  \includegraphics[width=1.0\columnwidth] {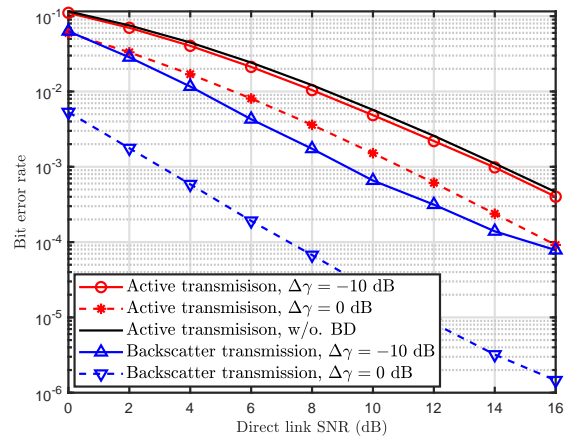}
  \caption{\small BER performance for the active transmission and the backscatter transmission in the CBC: Each BD symbol period covers $20$ active transmitter symbol periods and $\Delta \gamma$ is the average power ratio of the backscatter link to the active link.}
  \label{fig:sim1}
\end{figure}

Obligate SC describes a radio ecosystem in which one SR heavily relies on the others, and cannot realize its communication objective by itself. One of the frequently cited instance is cognitive backscatter communication (CBC) system, which achieves mutual benefits innately~\cite{yang2018cooperative,liang2020symbiotic}. Specifically, as Fig. \ref{fig:CBC} shows, the CBC system consists of two communication systems: the active transmission and the backscatter transmission systems. The active transmission system shares its spectrum and energy resources with the backscatter device (BD), and enables it to perform backscatter transmission via reflecting the active signal by periodically varying the reflecting coefficients. Meanwhile, the BD reflects the incoming active signal to the receiver and provides additional signal power source to the active transmission. In this symbiosis, the backscatter transmission system obtains the transmission opportunity without requiring additional spectrum and infrastructures, and in return the active transmission system has the multi-path diversity gain, thereby yielding mutual benefits. In Fig~\ref{fig:sim1}, it is shown that the active transmission achieves better bit-error-rate (BER) performance through the symbiosis with the backscatter transmission, and meanwhile enables the backscatter transmission.

The symbiosis in the CBC forms a resource-resource mutualism, similar to the legume-rhizobium mutualism. In particular, the active transmission provides its spectrum and energy resources (like nutritional materials to rhizobium) to the backscatter transmission, while the BD collects the primary signal energy resource (like nitrogen to legume plants) in the air for the active transmission.

Another example of obligate SC is symbiotic relaying shown in Fig. \ref{fig:CBC}, which consists of two SRs, the primary communication system and the secondary system~\cite{nadkar2011cross}. The primary communication has adequate spectrum, but its transmission link is unsustained and weak due to the obstacle block. The secondary system does not have radio spectrum, thus relies on the primary system to transmit. Thus, the primary system seeks help from the secondary system to provide high-quality relaying service. In return, after relaying the primary transmission, the secondary system is allowed to occupy the primary spectrum resource to transmit. By this way, the two systems achieve mutual benefits from the cooperation.
In this symbiosis, the primary communication system supplies the precious spectrum (like nectar to bees) and attract the secondary system to provide high quality relaying service (carrying pollen from flower to flower), which forms a resource-service mutualism, similar to the plant-pollinator symbiosis in biology.

\subsection{Facultative SC}

Facultative SC refers to a radio ecosystem in which the SRs are relatively independent and each of them can still perform its functions without the others. Nevertheless, each SR can benefit from the symbiosis established in the radio ecosystem. One typical example of facultative SC is the symbiosis of WiFi and licensed-assisted access LTE (LAA-LTE) in unlicensed band~\cite{laalte2019Tan}. As shown in Fig.~\ref{fig:CBC}, the LAA-LTE and WiFi systems share the same unlicensed spectrum but use different transmission mechanisms. Specifically, the WiFi system adopts the carrier sense multiple access with collision avoidance (CSMA/CA) mechanism, while the LAA-LTE system utilizes the central scheduling mechanism to access the channels.

Traditionally, WiFi and LAA-LTE systems operate individually based on their own transmission protocols. However, it is envisioned that Wi-Fi and LAA-LTE can achieve complementary benefits from the symbiotic relationships through resource/service exchange. On one hand, the LAA-LTE can serve some WiFi users and provide them with controllable access and guaranteed quality-of-service. In return, for the WiFi system, offloading some users to the well-managed LAA-LTE system will reduce the collision probability thus improving the performance of the users remained in the WiFi system.
The symbiosis between the LAA-LTE and WiFi systems forms a resource-service mutualism, where the WiFi system shares its spectrum resource with the LAA-LTE, and the LAA-LTE in return provides offloading services to some of the WiFi users.

\section{Symbiotic Coevolution}\label{sec:exchange}

For facultative SC, the SRs can work individually, thus the symbiotic relationship can be achieved through symbiotic coevolution, the core of which is evolutionary cycle.

\begin{figure}
  \centering
  \includegraphics[width=0.99\columnwidth] {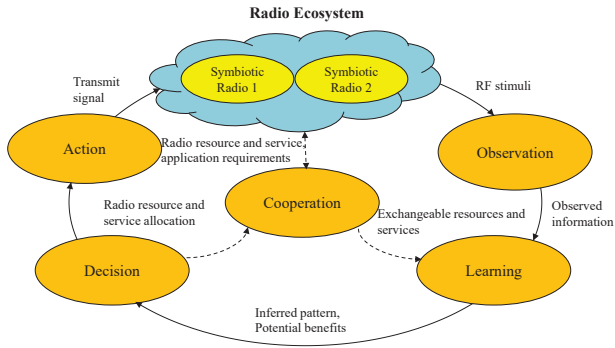}
  \caption{\small{Evolutionary cycle for an SR in SC.}}
  \label{fig:cycle}
\end{figure}

\subsection{Evolutionary Cycle}

Self-evolution has been traditionally applied to individual radio system while treating the others as part of the environment. SC breaks the boundary of the individual SRs, and enables the SRs to coevolve into mutualistic symbiosis. To achieve this, the coevolution mechanism should be introduced into each SR, and incites the SR to share its radio resources and services with others. An evolutionary cycle is proposed to empower the SRs to support the coevolution process. As shown in Fig. \ref{fig:cycle}, the key components in each cycle include cooperation, observation, learning, decision, and action.

The evolutionary cycle starts with the cooperation among the SRs, where each SR informs the others its state information, including the radio resources and services that it can provide, the QoS requirements for its application, and the benefit it obtains so far.
Then, the SR observes the RF stimuli to implicitly acquire the state in the radio environment. Together with the state information of other SRs obtained from the cooperation, each SR can learn and infer the pattern of the environment and the potential benefits from the other SRs. Based on the inferred information, each SR makes its adaptive decision on the radio resource and service allocation, such as spectrum management, power control, rate adaptation, access strategy and so on. Finally, each SR takes action accordingly, shares its radio resources and provides its radio services via the cooperation. By passing through the evolutionary cycle round and round, the SC intelligently reassigns the radio resources and services to the SRs. Thus, the SRs in SC may have additional degree-of-freedom to fulfill the stringent requirements, yielding the SRs to coevolve into mutualistic partners.

Take the LAA-LTE as an example, to achieve mutualism between the WiFi and LAA-LTE systems, LAA-LTE users occupy the channel for some of the slots in each frame and the WiFi users utilize the remaining part of the frame. If the occupied slots are too short for LAA-LTE within a frame, the channel may be idle for a long time when the WiFi system accomplishes the transmission, leading to low spectrum utilization. On the other hand, if the occupied slots are too long, the WiFi system is difficult to finish the transmission successfully, resulting in a poor performance for WiFi transmission. As a result, the evolutionary cycle can be introduced to improve the spectrum usage and simultaneously protect the WiFi system. Specifically, the LAA-LTE system continuously observes the WiFi activities and intelligently learns the WiFi traffic models. At the same time, the WiFi system may cooperatively inform its traffic demand to the LAA-LTE system. With the observed and informed information, the LAA-LTE system makes decisions about the transmission time and takes action accordingly. With this adaptive action policy, the spectrum efficiency can be improved and the WiFi system can be protected, yielding mutualistic symbiosis.


\subsection{Multi-Agent Learning}

Through the evolutionary process, the SRs in SC coevolve into mutualistic symbiosis by exchanging the radio resource and service, thus the SC can be regarded as a multi-agent system. As an intelligent agent, each SR interacts with each other and makes decisions in a distributed manner, thus multi-agent reinforcement learning can be used to execute the distributed learning.


In multi-agent reinforcement learning, each SR learns its own best policy individually through interacting with other SRs and its environment over time via trial and error \cite{bucsoniu2010multi}. For example, in the symbiosis of WiFi and LAA-LTE, the LAA-LTE system iteratively observes the current environment state (WiFi activities), selects an action (transmission time) according to the current policy, and then gets an immediate reward and a new environmental state. The reward function for LAA-LTE or for WiFi can correspond to different performance metrics, e.g., capacity or throughput.
Each system makes individual decisions based on its local knowledge and the received state information of other users. The collective objective of all users is to maximize the globally averaged benefits, e.g., spectrum efficiency. To achieve this, each user adjusts its policy repeatedly to approach the optimal policy by maximizing the long-term reward, which consists of the immediate reward and future reward. In general, transition probability from the current state to the next state is used to calculate the future reward. However, the LAA-LTE system is difficult to obtain the transition probability, especially in a complicated environment. In that case, a model-free RL algorithm is widely used, such as Q-learning algorithm, which constructs a lookup Q-table to indicate the long-term rewards of all possible state-action pairs. When the size of the state and action spaces is large, it is difficult to store the large Q-table and the performance of the Q-learning algorithm is degraded. Alternatively, deep RL can be used to make an optimal policy by introducing a deep neural network to estimate the long-term reward.

Multi-agent learning makes the SC system smarter and less complex. On one hand, the SR can use local and outdated information to extract the hidden pattern and make decisions accordingly, thereby reducing the system complexity and overhead. On the other hand, multi-agent learning is applied to all SRs within the SC, endowing the SC system with intelligence and making the symbiotic coevolution achievable.

\section{Symbiotic Synthesis}\label{sec:MOO}

As aforementioned, the symbiotic relationship in SC can also be artificially constructed via the symbiotic synthesis, which involves co-designing the SRs for collective objectives. Multiple-objective optimization (MOO) can be used to model the symbiotic synthesis, in conjunction with objective function and constraint refinement introduced by resource and service exchange.

\subsection{Multiple-Objective Optimization}

In SC, each SR has its own performance metrics and radio resource constraints. When MOO is used to model the symbiotic synthesis, the exchange of radio resources typically does not affect the objective functions of each SR, but expands the constraints on the variables, thus affects feasible region of the solution. The exchange of radio services, however, may change the way that the SR achieves its objective. For instance, the symbiotic relaying system provides the high-quality relaying service, thereby changing the function of the achievable rate for the primary transmission. Nevertheless, the attributes of radio resources and services in the MOO model are not limited to the above analysis but depend on the specific scenarios.

In MOO, a set of optimal points are generally solved instead of a single global solution. Each optimal point is called \emph{Pareto optimality}, which refers to a situation where no sub-objective can become better off without degrading at least one of the other sub-objectives.
In SC, the set of Pareto optimality points indicates all potential benefits that each SR can achieve with various resource and service exchange strategies, thereby guiding the design of the symbiotic synthesis according to the expected benefits for each SR.

Conventionally, the MOO problem can be solved by scalarization approach or Pareto approach. For the first method, the MOO problem is solved by scalarizing the objective function via the weighted sum scalarization or Chebychev scalarization \cite{marler2004survey}. The weighted values represent the contributions from the sub-objectives and are prior chosen before the optimization process.
With different weighted values for scalarization, different Pareto optimal solutions are generated. On the other hand, the Pareto approach aims to produce all the Pareto optimal solutions by exploration and Pareto filtering, multi-objective genetic algorithms, or weighted sum approach using weighted value scanning.

\begin{figure}
  \centering
  \includegraphics[width=1.0\columnwidth] {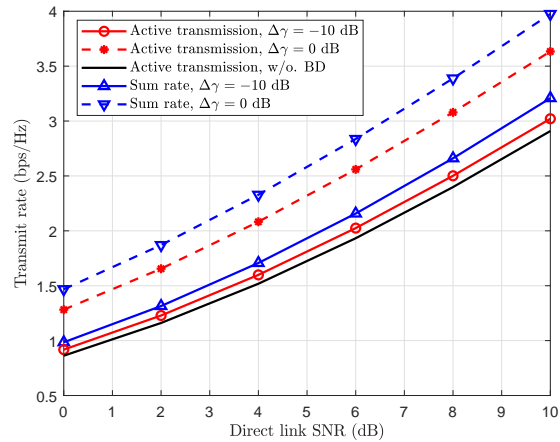}
  \caption{\small Transmit rate performance of CBC using sum rate optimization: Each BD symbol period covers $20$ active transmit symbol periods and $\Delta \gamma$ is the average power ratio of the backscatter link to the active link.}
  \label{fig:sim2} 
\end{figure}

\subsection{MOO for CBC}

Here, we use CBC as an example to show how to achieve symbiotic synthesis between the active and backscatter transmissions. In the CBC, we formulate a MOO problem which aims to optimize the active transmit rate and the backscatter transmit rate. In particular, we exploit the weighted sum rate method to achieve the symbiotic synthesis by designing the symbol period of the backscatter transmission, when the symbol period of the active transmission is given~\cite{long2019symbiotic}. Note when the BD symbol period is sufficiently large and covers a sequence of consecutive active transmit symbols, the backscatter transmission provides the additional multi-path signal rather than interference to the active transmission. On the other hand, the backscatter transmission is enabled by the active transmission. As shown in Fig.~\ref{fig:sim2}, this symbiotic synthesis helps to enhance the active transmission and increase the sum spectrum efficiency with the introduction of the backscatter transmission.

Therefore, by solving the MOO problem, the CBC ingeniously optimizes the key operating parameters and achieves mutualistic symbiosis between the active and backscatter transmissions.

\section{Discussions} \label{sec:chall}

SC is a new wireless communication paradigm which improves the utilization efficiency of the radio resources, and enhances the over performance of the SRs. As SC involves interdiciplinary study related to wireless communication, evolutionary theory, and artificial intelligence, to make SC a reality, there are still some technical challenges and open problems which need to be addressed.

\subsection{System Architectures}

The exchange of the radio resource and services among SRs calls for novel cross-layer design for SC. Also, a common control channel is needed in SC to support cooperations among the SRs. In addition, the execution of evolutionary cycle involves intensive computations and communication overhead. Thus novel system architecture with the joint consideration of communication and computing deserves considerable investigations.

\subsection{Symbiotic Learning and Evolution}
Symbiotic learning, which incorporates the benefit states of the SRs into the learning process can be a potential solution to achieve evolutionary symbiosis~\cite{eguchi2006study}. However, the exchange of radio resources and services could potentially change the objective functions, thus the rewards of the agents. Such property may affect the convergence and stability of the multi-agent learning, and the research along this direction is needed.

\subsection{Applications in 6G Networks}

It has been shown that the coexistence of cellular and IoT networks may achieve mutual benefits through adopting CBC~\cite{liang2020symbiotic}. This offers us new solutions to design ultra-massive access for the 6th generation (6G) networks. It is also foreseen that in future space-air-ground networks, radar-communications integrated networks, and emergency-cellular networks, the individual network may benefit from the SC paradigm through intelligent resource and service exchange.

\subsection{Economic Aspects and Incentive Mechanisms}

Compared to conventional CR which supports dynamic spectrum sharing in a non-avoidable interfering basis, SC advocates purposely-designed spectrum sharing achieving mutualistic benefits, thus providing us new guidelines to design radio spectrum policies for future wireless communication systems. To promote the adoption of such paradigm, however, it is required to investigate the economic aspects of SC, and to design suitable incentive mechanisms to encourage the SRs to participate in SC.

\subsection{Concluding Remarks}

Mutualistic symbiosis is the always the best of the world. While this paper addresses the symbiosis in radio space, the concept of symbiosis has recently been applied to the field of computing to achieve human-AI symbiosis\footnote{https://sloanreview.mit.edu/article/creating-the-symbiotic-ai-workforce-of-the-future/}. It is foreseen that there will be more and more applications of symbiosis in various fields, and we are expecting a symbiotic world in the future.


\bibliographystyle{IEEEtran}
\bibliography{ref_AmBC}

\end{document}

%% file: preamble3.tex
\usepackage{cite}
\usepackage[cmex10]{amsmath}
\usepackage{amssymb}
\usepackage{amsthm}
\usepackage{mathrsfs}
\usepackage{bm}
\usepackage[mathscr]{eucal}
\usepackage{amssymb,amsmath,amsthm,amsfonts,latexsym}
\usepackage{amsmath,graphicx,bm,xcolor,url}
\usepackage{graphicx}
\usepackage{latexsym}
\usepackage{CJK}
\usepackage{indentfirst}
\usepackage{geometry}
\usepackage{psfrag}
\usepackage{setspace}
\usepackage{algorithmic}
\usepackage{algorithmic, cite}
\usepackage{algorithm}
\usepackage{array}
\usepackage{mdwmath}
\usepackage{mdwtab}
\usepackage{eqparbox}
\usepackage{url}
\usepackage{epstopdf}
\usepackage{epsfig,epsf,psfrag}
\usepackage{fixltx2e}
\usepackage{verbatim}
\usepackage{textcomp}
\usepackage{psfrag} 

\usepackage{subcaption}

\theoremstyle{plain}

\theoremstyle{remark}

\theoremstyle{plain}

\theoremstyle{remark}

\theoremstyle{plain}

\theoremstyle{remark}

\theoremstyle{remark}

\theoremstyle{remark}

\theoremstyle{remark}

\theoremstyle{remark}

\theoremstyle{remark}

\catcode`~=11 \def\UrlSpecials{\do\~{\kern -.15em\lower .7ex\hbox{~}\kern .04em}} \catcode`~=13

\allowdisplaybreaks[4]

\DeclareMathAlphabet{\mathbsf}{OT1}{cmss}{bx}{n}
\DeclareMathAlphabet{\mathssf}{OT1}{cmss}{m}{sl}

\DeclareSymbolFont{bsfletters}{OT1}{cmss}{bx}{n}
\DeclareSymbolFont{ssfletters}{OT1}{cmss}{m}{n}
\DeclareMathSymbol{\bsfGamma}{0}{bsfletters}{'000}
\DeclareMathSymbol{\ssfGamma}{0}{ssfletters}{'000}
\DeclareMathSymbol{\bsfDelta}{0}{bsfletters}{'001}
\DeclareMathSymbol{\ssfDelta}{0}{ssfletters}{'001}
\DeclareMathSymbol{\bsfTheta}{0}{bsfletters}{'002}
\DeclareMathSymbol{\ssfTheta}{0}{ssfletters}{'002}
\DeclareMathSymbol{\bsfLambda}{0}{bsfletters}{'003}
\DeclareMathSymbol{\ssfLambda}{0}{ssfletters}{'003}
\DeclareMathSymbol{\bsfXi}{0}{bsfletters}{'004}
\DeclareMathSymbol{\ssfXi}{0}{ssfletters}{'004}
\DeclareMathSymbol{\bsfPi}{0}{bsfletters}{'005}
\DeclareMathSymbol{\ssfPi}{0}{ssfletters}{'005}
\DeclareMathSymbol{\bsfSigma}{0}{bsfletters}{'006}
\DeclareMathSymbol{\ssfSigma}{0}{ssfletters}{'006}
\DeclareMathSymbol{\bsfUpsilon}{0}{bsfletters}{'007}
\DeclareMathSymbol{\ssfUpsilon}{0}{ssfletters}{'007}
\DeclareMathSymbol{\bsfPhi}{0}{bsfletters}{'010}
\DeclareMathSymbol{\ssfPhi}{0}{ssfletters}{'010}
\DeclareMathSymbol{\bsfPsi}{0}{bsfletters}{'011}
\DeclareMathSymbol{\ssfPsi}{0}{ssfletters}{'011}
\DeclareMathSymbol{\bsfOmega}{0}{bsfletters}{'012}
\DeclareMathSymbol{\ssfOmega}{0}{ssfletters}{'012}

\def\norm#1{\left\| #1 \right\|}
\def\norm2#1{\left\| #1 \right\|_2}
\def\norm22#1{\left\| #1 \right\|_2^2}

\newcommand{\qednew}{\nobreak \ifvmode \relax \else
      \ifdim\lastskip<1.5em \hskip-\lastskip
      \hskip1.5em plus0em minus0.5em \fi \nobreak
      \vrule height0.75em width0.5em depth0.25em\fi}